\documentclass{acm_proc_article-sp}

\usepackage{algorithm}
\usepackage{algorithmic}

\usepackage{listings}
\begin{document}

\title{Model-based optimization of MPDATA on Intel Xeon Phi through load imbalancing}

\numberofauthors{3} 

\author{
\alignauthor
Alexey Lastovetsky 
\\
       \affaddr{University College Dublin}\\
       \affaddr{Belfield, Dublin 4, Irleand}\\
       \email{alexey.lastovetsky@ucd.ie}
\alignauthor
Lukasz Szustak 
\\
       \affaddr{Czestochowa University of Technology}\\
       \affaddr{Dabrowskiego 69, 42-201 Czestochowa, Poland}\\
       \email{lszustak@icis.pcz.pl}       
\alignauthor
Roman Wyrzykowski 
       \affaddr{Czestochowa University of Technology}\\
       \affaddr{Dabrowskiego 69, 42-201 Czestochowa, Poland}\\
       \email{roman@icis.pcz.pl}   
}

\maketitle
\begin{abstract}
Load balancing is a widely accepted technique for performance optimization of scientific applications on parallel architectures. Indeed, balanced applications do not waste processor cycles on waiting at points of synchronization and data exchange, maximizing this way the utilization of processors. In this paper, we challenge the universality of the load-balancing approach to optimization of the performance of parallel applications. First, we formulate conditions that should be satisfied by the performance profile of an application in order for the application to achieve its best performance via load balancing. Then we use a real-life scientific application, MPDATA, to demonstrate that its performance profile on a modern parallel architecture, Intel Xeon Phi, significantly deviates from these conditions. Based on this observation, we propose a method of performance optimization of scientific applications through load imbalancing. We also propose an algorithm that finds the optimal, possibly imbalanced, configuration of a data parallel application on a set of homogeneous processors. This algorithm uses functional performance models of the application to find the partitioning that minimizes its computation time but not necessarily balances the load of the processors. We show how to apply this algorithm to optimization of MPDATA on Intel Xeon Phi. Experimental results demonstrate that the performance of this carefully optimized load-balanced application can be further improved by 15\% using the proposed load-imbalancing optimization.
\end{abstract}


\section{Introduction}
\label{sec:intro}
Load balancing is a widely accepted technique for optimization of the computation performance of scientific applications on parallel architectures. Indeed, the intuition suggests that unlike unbalanced applications, the balanced ones do not waste processor cycles on waiting at points of synchronization and data exchange, maximizing this way the utilization of the processors. 

In this paper, we challenge the universality of the load-balancing approach to optimization of the computation performance of parallel applications.  First, we try to understand the limitations of the load-balancing approach. We formulate conditions that should be satisfied by the performance profile of an application in order for the application to achieve its best performance via load balancing.

Then we use a real-life scientific application to demonstrate that its performance profile on a modern parallel architecture does not satisfy these conditions.
The application we use implements the Multidimensional Positive Definite Advection Transport Algorithm (MPDATA), which is one of the major parts of the dynamic core of the EULAG geophysical model \cite{SMO06}.
EULAG (Eulerian/semi-Lagrangian fluid solver) is an established numerical model developed for simulating thermo-fluid flows across a wide range of scales and physical scenarios \cite{PIO12,SMO90}. In particular, it can be used in numerical weather prediction (NWP), simulation of urban flows, areas of turbulence, ocean currents, etc. This solver, originally developed for conventional HPC systems, is currently being re-written for modern HPC platforms. In particular, MPDATA has been recently re-written and optimized for execution on an Intel Xeon Phi coprocessor \cite{SZU13,SZU15}.

In our experiments, we observe significant deviations of the MPDATA performance profile from the conditions required for applicability of the load-balancing techniques. Based on this observation, we propose a general method of performance optimization of scientific applications through load imbalancing as well as an algorithm that finds the optimal, possibly imbalanced, configuration of a data parallel application on a set of homogeneous processors. This algorithm uses functional performance models of the application  \cite{twamley2005, ijhpca2007} to find the partitioning that minimizes its computation time but not necessarily balances the load of the processors. Finally, we apply this algorithm to optimization of MPDATA on Intel Xeon Phi. Experimental results demonstrate that the performance of this carefully optimized load-balanced application can be further improved by 15\% using the proposed load-imbalancing method.



The contributions of the work presented in the paper are as follows:
\begin{itemize}
\item Formulation of the conditions that should be satisfied to guarantee that load balancing will minimize the computation time of parallel application.
\item Building the performance profile of a real-life scientific application on a modern HPC platform and demonstration of its significant deviation from the conditions that guarantee that load balancing be a safe technique for performance optimization.
\item A new optimization method that uses the performance profile for optimization of the application through its imbalancing.
\item A partitioning algorithm finding the optimal and generally speaking uneven distribution of computations of an application between homogeneous processing units using its functional performance model.
\item Application of the proposed partitioning algorithm to optimization of MPDATA on Intel Xeon Phi, resulting in further acceleration of this carefully optimized load-balanced application by up to 15\%.
\end{itemize}

The rest of the paper is structured as follows. Section~\ref{sec:per} overviews load-balancing techniques and formulates the conditions when these techniques would minimize the computation time of parallel applications. Section~\ref{sec:opt} analyzes the performance profile of MPDATA on Intel Xeon Phi and introduces the new approach to minimization of the computation time through load imbalancing. Section~\ref{sec:model}  introduces a partitioning algorithm for (uneven) distribution of computations between homogeneous processors, minimizing the computation time of the application. Section~\ref{sec:app} applies this algorithm to optimization of MPDATA on Xeon Phi. Section~\ref{sec:result} presents experimental results, and Section~\ref{sec:con} concludes the paper.

\section{Load balancing and performance}
\label{sec:per}

In this section we overview load-balancing techniques used for optimization of the performance of parallel scientific applications on both homogeneous and heterogeneous platforms. We also formulate the conditions when application of these techniques will optimize the computation performance.

Load balancing is a widely accepted and commonly used approach to performance optimization of scientific applications on parallel architectures. Load balancing algorithms can be classified as static or dynamic. Static algorithms (for example, those based on data partitioning) \cite{Fatica2009, Yang2010, Ogata2008, ijhpca2007} require a priori information about the parallel application and platform. This information can be gathered either at compile-time or runtime. Static algorithms are also known as predicting-the-future because they rely on accurate performance models as input to predict the future execution of the application. Static algorithms are particularly useful for applications where data locality is important because they do not require data redistribution. However, these algorithms are unable to balance on non-dedicated platforms, where load changes with time. Dynamic algorithms (such as task scheduling and work stealing) \cite{Linderman2008, Augonnet2009, QuintanaOrti2009} balance the load by moving fine-grained tasks between processors during the calculation. Dynamic algorithms do not require a priori information about execution but may incur significant communication overhead due to data migration. Dynamic algorithms often use static partitioning for their initial step due to its provably near-optimal communication cost, bounded tiny load imbalance, and lesser scheduling overhead \cite{Song2012}. 

Whatever load balancing algorithm is used, the goal is always to minimize the computation time of the application. The intuition behind the assumption that balancing the application improves its performance is the following: a balanced application does not waste processor cycles on waiting at points of synchronization and data exchange, maximizing this way the utilization of the processors. Is this assumption always true? To answer this question, let us formulate the assumption in a mathematical form. Consider an application, the computation performance of which can be modeled by speed functions. Namely, let $p$ parallel processors be used to execute the application, and let $s_i(x)$ be the speed of execution of the workload of size $x$ by processor $i$. Here the speed can be measured in floating point operations per second or any other fix-sized computation units per unit time. The size of workload can be characterized by the problem size (say, the number of cells in the computational domain or the matrix size) or just by the number of equal-sized computational units. Anyway, the speed $s_i(x)$ is calculated as $\frac{x}{t_i(x)}$, where $t_i(x)$ is the execution time of the workload of size $x$ on processor $i$. Using these definitions, we can formulate the following theorem.

\emph{Theorem 1}:
\label{theo:speed} 
Let $s_i(x)>0$ ($x>0$) be the speed of processor ${i \in \{1,\ldots,p\} }$, and $\forall{\Delta x>0} \colon$ $\frac{s_i(x)} {x}$ $\geq$ $\frac{s_i (x+\Delta x)} {x+\Delta x}$. Let $x_1+\ldots+x_p=n>0$ and $\frac{s_1(x_1)} {x_1}$ $=\ldots=$ $\frac{s_p(x_p)} {x_p}$. Then, $\forall{y_1,\ldots,y_p}$ $>$ $0$ such that $(y_1,\ldots,y_p)$ $\neq$ $(x_1,\ldots,x_p)$ and $y_1+\ldots+y_p=n \colon$ $\max_{i}\frac{y_i}{s_i(y_i)}$ $\geq$ $\frac{x_1}{s_1(x_1)}$.

\emph{Proof}: 
As  $(y_1,\ldots,y_p)$ $\neq$ $(x_1,\ldots,x_p)$ and $y_1+\ldots+y_p=x_1+\ldots+x_p$, then there exists ${k \in \{1,\ldots,p\} }$ such that $y_k > x_k$. Therefore, $\max_{i}\frac{y_i}{s_i(y_i)}$ $\geq$ $\frac{y_k}{s_k(y_k)}$ $=$ $\frac{x_k+(y_k-x_k)}{s_k(x_k+(y_k-x_k))}$ $\geq$ $\frac{x_k}{s_k(x_k)}$ $=$ $\frac{x_1}{s_1(x_1)}$.\\

Theorem 1 states that in order to guarantee that the balanced configuration of the application will execute the workload of size $n$ faster than any unbalanced configuration, the speed functions  $s_i(x)$, characterizing the performance profiles of the processors, should satisfy the condition:
\begin{equation}
  \forall{\Delta x>0} \colon \frac{s_i(x)} {x} \geq \frac{s_i (x+\Delta x)} {x+\Delta x}
\label{eq:cond}
\end{equation}

\begin{figure}[h]
\begin{center}
\includegraphics[width=0.5\textwidth]{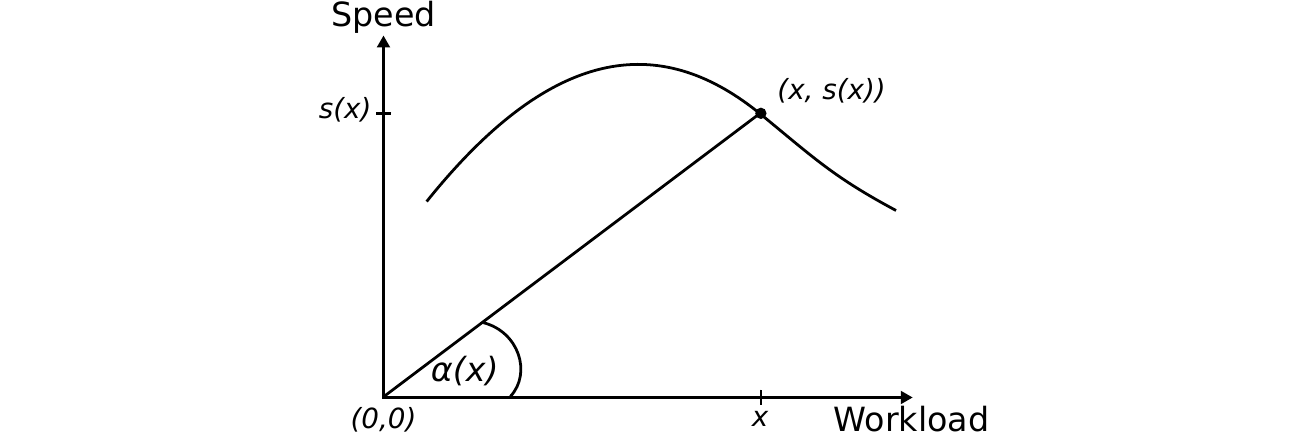}
\caption{Example of speed function suitable for minimization of computation time through load balancing. Angle $\alpha(x)$ represents the computation time: the greater the angle, the shorter the computation time.}
\label{fig:graph0}
\end{center}
\end{figure}

Geometrically, it can be illustrated as follows. If we plot a speed function as shown in Figure~\ref{fig:graph0}, then the angle $\alpha (x)$ between the straight line, connecting the point $(0,0)$ and the point $(x, s(x))$ on the speed curve, and the $x$-axis will be inversely proportional to the execution time of the workload of size $x$ by the processor. Indeed, the cotangent of this angle is directly proportional to the ratio $\frac{x}{s(x)}$ representing the execution time of the workload $x$. Therefore, larger angles correspond to shorter execution times. The condition~\ref{eq:cond} means that the increase of the workload, $x$, will never result in the decrease of the execution time, or equivalently in the increase of the angle $\alpha (x)$.

The main body of the load balancing algorithms designed for performance optimization explicitly or implicitly assume that the speed of processor does not depend on  the size of workload \cite{Fatica2009, Yang2010, Luk2009, cierniak, kalinov2001, plaza2011}. In other words, the speed functions $s_i(x)$  are assumed to be positive constants, in which case the condition~\ref{eq:cond} is trivially satisfied. More advanced algorithms are based on functional performance models (FPMs), which represent the speed of processor by a continuous function of the problem size \cite{twamley2005, ipdps2004}. However, the shape of the function is not  arbitrary but has to satisfy the following assumption \cite{ijhpca2007}: Along each of the problem size variables, either the function is monotonically decreasing, or there exists point $x$ such that
\begin{itemize}
\item On the interval $[0, x]$, the function is
\begin{itemize}
\item monotonically increasing,
\item concave, and
\item any straight line coming through the origin of the coordinate system intersects the graph of the function in no more than one point.
\end{itemize}
\item On the interval $[x, \infty)$, the function is monotonically decreasing.
\end{itemize}
These restrictions on the shape of speed functions guarantee that the efficient load balancing algorithms, proposed in \cite{Ilic2011,Colaco,lastovetsky2007data, clarke2011dynamic, clarke2012column,alonazi2015}, will always return a unique solution, minimizing the computation time. At the same time, it is easy to show that the restrictions imposed on FPMs will make them comfortably satisfy the condition~\ref{eq:cond}. 

Thus, the state-of-the-art load balancing algorithms designed for optimization of the computation performance of parallel applications assume that their performance profiles satisfy the condition~\ref{eq:cond}.  Therefore, correct application of such algorithms requires that the experimental speed points be approximated by a function satisfying this condition. This approximation step may significantly distort the actual performance profile and lead to a substantially non-optimal solution. 

\section{Optimization of parallel applications through load imbalancing}
\label{sec:opt}
In this section, we demonstrate that the performance profile of real-life scientific applications on modern parallel platforms may significantly deviate from the conditions, which guarantee that load balancing will always optimize their computation performance. Based on this observation, we propose an optimization method that uses the performance profile for optimization of the application through its imbalancing.

\begin{figure}[h]
\begin{center}
\includegraphics[width=0.5\textwidth]{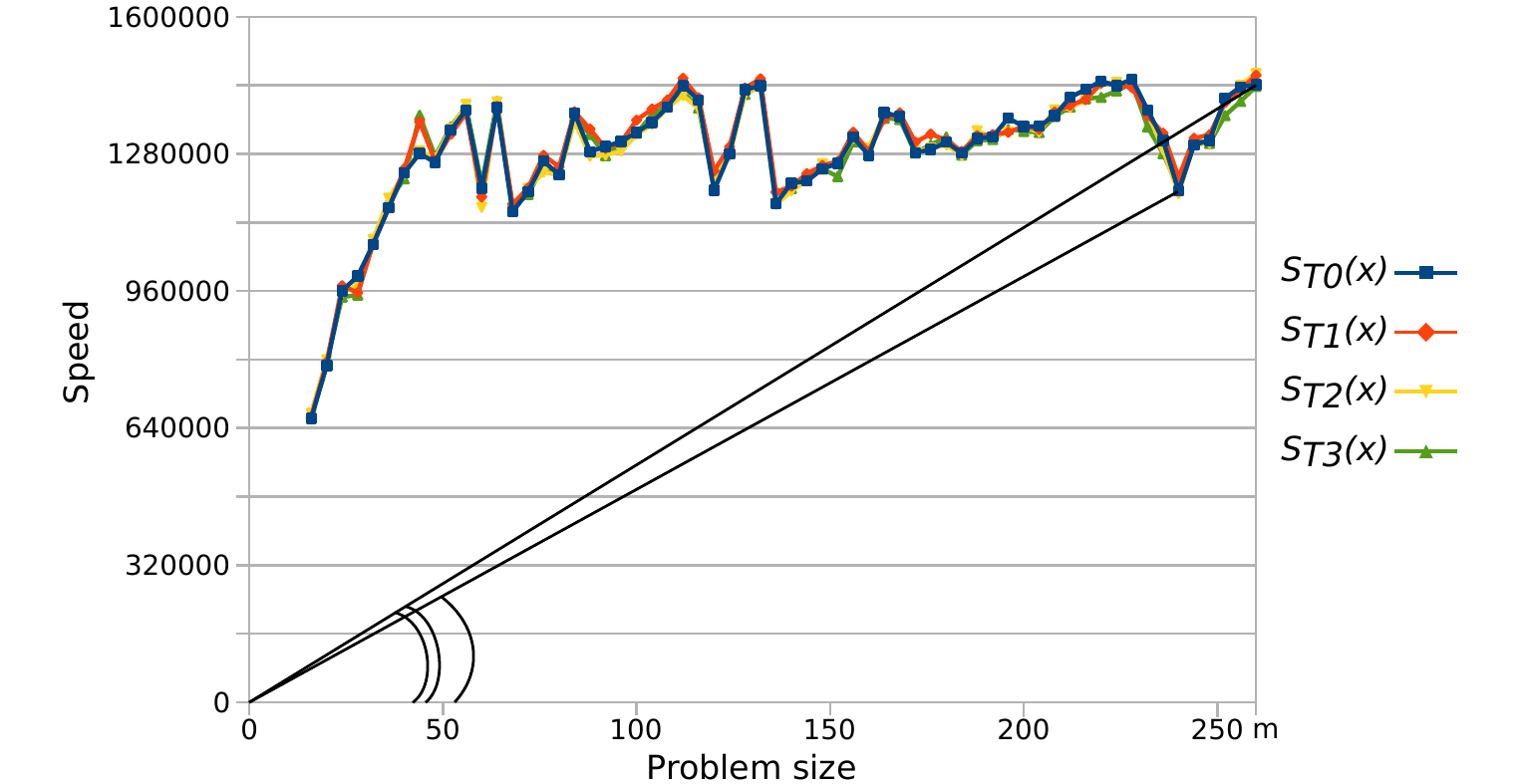}
\caption{ Speed functions of Intel MIC built for four 15-core teams, each processing in parallel a $120\times m\times 128$ sub-domain. The speed is measured in cells per second, while the problem size is represented by $m$. }
\label{fig:graph1}
\end{center}
\end{figure}

In this work, we build the performance profile of MPDATA on Intel Xeon Phi. MPDATA is a core component of EULAG (Eulerian/semi-Lagrangian fluid solver), which is an established computational model developed for simulating thermo-fluid flows across a wide range of scales and physical scenarios. Its carefully optimized data-parallel implementation on a 60-core Intel Xeon Phi \cite{SZU15}  partitions the 3D rectilinear $n\times n\times l$ domain into four equal $\frac{n}{2}\times \frac{n}{2} \times l$ sub-domains, each allocated to a team of 15 cores. This configuration of the application is the best load-balanced configuration identified in \cite{SZU15}.  The experimentally constructed speed functions of these four teams, each processing (in parallel) a $120\times m\times 128$ sub-domain, are shown in Figure~\ref{fig:graph1}.

This graph clearly shows that for many $m$ and $\Delta m$ the speed of processing of the $120\times m\times 128$ sub-domain will be significantly lower than the speed of processing of the $120\times (m+\Delta m)\times 128$ sub-domain. Moreover, we can also see that $\alpha(m+\Delta m) >\alpha(m)$ for some such $m$ and $\Delta m$, which means that the time of processing of the $120\times m\times 128$ sub-domain will be longer than the time of processing of the $120\times (m+\Delta m)\times 128$ sub-domain. The latter observation can be used to speed up the execution of the application. Namely, if we re-partition the equally partitioned domain so that two teams get  $120\times (m+\Delta m)\times 128$ sub-domains and two other teams get  $120\times (m-\Delta m)\times 128$ sub-domains, and $\min\{\alpha(m+\Delta m), \alpha(m-\Delta m)\}>\alpha(m)$, than this unequal and unbalanced partitioning will result in faster execution.

In general, if the performance profile of an application violates the condition~\ref{eq:cond}, that is,
\begin{equation}
 \exists {i \in \{1,\ldots,p\}} \exists{x>0} \exists{\Delta x>0}\colon \frac{s_i(x)} {x} < \frac{s_i (x+\Delta x)} {x+\Delta x}
\label{eq:not-cond}
\end{equation}
and the balanced configuration of the application allocates the workload of size $x$ to processor $i$, then the application can be accelerated if we reduce the accumulated workload of all processors but processor $i$ by $\Delta x$ so that none of these processors would increase its execution time, and allocate this additional workload to processor $i$.  This  method can be applied to optimization of parallel applications on both heterogeneous and homogeneous platforms.

\section{Model-based partitioning algorithm for optimal load imbalancing}
\label{sec:model}
In this section, we develop the proposed approach for a relatively simple case and introduce a partitioning algorithm that finds the optimal distribution of computations of an application between homogeneous processors using the functional performance model of the application.

Consider the following problem. Let $p$ identical parallel processors be used to execute the workload of size $n$, and let $s(x)$ be the speed of execution of the workload of size $x$ by a processor. Let $\Delta x$ be the minimal granularity of workload so that each processor can be only allocated a multiple of $\Delta x$. The problem is to find the distribution of the workload of size $n$ between the $p$ processors, which minimizes the computation time of its parallel execution.

For simplicity, we assume that $\frac{n}{p}$ be a multiple of $\Delta x$ and $p$ be an even number. Then the following Algorithm~\ref{Alg:Homo} will solve this problem.

\begin{algorithm}
\caption{Optimal distribution of workload between homogeneous processors}
\begin{algorithmic}
\STATE $x_{ropt}=x_{lopt}=x_r=x_l=\frac{n}{p}$
\STATE $t_{min}=\frac{\frac{n}{p}}{s(\frac{n}{p})}$
\REPEAT
\STATE $x_r+=\Delta x$
\STATE $x_l-=\Delta x$
\STATE $t=\max(\frac{x_r}{s(x_r)},\frac{x_l}{s(x_l)})$
\IF{$t<t_{min}$}
\STATE $t_{min}=t$
\STATE $x_{ropt}=x_r$
\STATE $x_{lopt}=x_l$
\ENDIF
\UNTIL{$x_r<n$}
\end{algorithmic}
\label{Alg:Homo}
\end{algorithm}

This algorithm returns the optimal distribution of the workload of size $n$  between $p$ processors so that each odd processor receives the workload of size $x_{lopt}$, and each even processor receives the workload of size $x_{ropt}$. It can be proved that the solution returned by this algorithm will always minimize the execution time of the given workload $n$. Note that the traditional load-balanced approach would assign the workload of size $\frac{n}{p}$ to all processors.

It is obvious that if we replace the speed function $s(x)$ by any function $a\times s(x)$, where $a=const$, then this algorithm  will return the same solution. We will use this property when applying Algorithm~\ref{Alg:Homo} in Section~\ref{subsec:model}.

Algorithm~\ref{Alg:Homo} can be easily generalized for an arbitrary (not only even) number of processors.

\section{Application}
\label{sec:app}
In this section, we apply the partitioning algorithm proposed in Section~\ref{sec:model} to optimization of MPDATA on Intel Xeon Phi.

\subsection{Intel MIC overview}

The Intel MIC architecture is a relatively new system for high performance computing \cite{MICA}.
Intel MIC combines many integrated Intel CPU cores into a single chip.
This architecture is built to provide a general-purpose programming environment similar to that provided for Intel CPUs.
It is capable of running applications written in industry-standard programming languages such as Fortran, C, and C++.
The Intel Xeon Phi (codenamed Knights Corner) is the first product based on Intel MIC architecture.
This coprocessor is delivered in form factor of a PCI express device, and can not be used as a stand-alone processor.
However, it allows users to directly run individual applications in the native mode without the support of CPU.

In this study, we use the top-of-the-line Intel Xeon Phi 7120P coprocessor.
It contains 61 cores clocked at 1.238~GHz, and 16~GB of on-board memory.
As the Intel MIC architecture supports four-way hyper-threading, it totally gives 244 logical cores (threads) for a single chip.
This coprocessor provides 352~GB/s of memory bandwidth.
An important component of each Intel Xeon Phi processing core is its vector processing unit (VPU) \cite{SZU15}, that significantly increases the computing power.
Each VPU supports a new 512-bit SIMD instruction set called Intel Initial ManyCore Instructions.   
The theoretical peak performance of Intel Xeon Phi~7120P is 1208~GFlop/s for double precision numbers.

\subsection{Introduction to MPDATA}

The MPDATA algorithm is a general approach for integrating the conservation laws of geophysical fluids on micro-to-planetary scales \cite{SMO01}.
It belongs to the class of methods for the numerical simulation of fluid flows which are based on the sign-preserving properties of upstream differencing.
The MPDATA scheme allows for solving advection problems, and offers several options to model a wide range of complex geophysical flows.

MPDATA corresponds to the group of nonoscillatory forward-in-time algorithms.
The number of required time steps depends on a type of simulated physical phenomenon, and can exceed few millions especially when considering the MPDATA algorithm as a part of the EULAG model. 
For detailed description of the MPDATA mathematical scheme, the reader is referred to \cite{SMO06,SMO90,SMO01}.

Each MPDATA time step is determined by a set of 17 computational stages, where each stage is responsible for calculating elements of a certain matrix.
These stages represent stencil codes which update grid elements according to different patterns. Listing~\ref{lst:MPDATAsrc} shows a part of the 3D~MPDATA stencil-based implementation for the 8-th stage.

\begin{lstlisting}[caption={Part of 3D MPDATA stencil-based implementation}, label={lst:MPDATAsrc}]
/*...*/
//stage 8
for( ... ) // i - dimension
  for( ... ) // j - dimension
    for( ... ) // k - dimension
      mx[i,j,k]=max(x[i][j][k],
                x[i-1][j][k], x[i+1][j][k],
                x[i][j-1][k], x[i][j+1][k],
                x[i][j][k-1], x[i][j][k+1]);
/*...*/
\end{lstlisting}

The stages are dependent on each other: outcomes of prior stages are usually input data for the subsequent computations.
Every stage reads a required set of matrices from the main memory, and writes results to the main memory after computation.
In consequence, a significant traffic to the main memory is generated, which mostly limits the performance of novel architectures.
A single MPDATA time step requires 5 input matrices, and returns one output matrix that is necessary for the next step.

\subsection{Adaptation of 3D MPDATA to Intel Xeon Phi coprocessor}
\label{subsec:adaptation}

In our previous work \cite{SZU13,SZU15}, we proposed the adaptation of 3D MPDATA to Intel Xeon Phi coprocessors.
The proposed decomposition contributes to ease the memory and communication bounds, and to better exploit computation resources of Intel Xeon Phi.
The resulting adaptation is based on the following methodology:
\begin{itemize}
\item (3+1)D decomposition of MPDATA heterogeneous stencil computations;
\item partitioning of threads into independent work teams;
\item parallelization of MPDATA computations;
\item scheduling for multicore and manycore systems.
\end{itemize}

To workaround the memory-bound nature of MPDATA, we proposed the (3+1)D decomposition of MPDATA stencil computation \cite{SZU15}.
The main aim of this decomposition is to take advantage of cache memory reuse by transferring the data traffic associated with all intermediate computation from the main memory to the cache hierarchy.
This aim is achieved by using a combination of two \mbox{well-known} loop optimization techniques: loop tiling and loop fusion.
Such an approach allows us to reduce the main memory traffic at the cost of additional computations associated with 
extra areas (\emph{halo}) of all intermediate matrices.
Another advantage of this approach is the possibility of reducing the main memory consumption because all intermediate results are stored only in the cache memory.

\begin{figure}[h!]
\begin{center}
\includegraphics[width=0.5\textwidth]{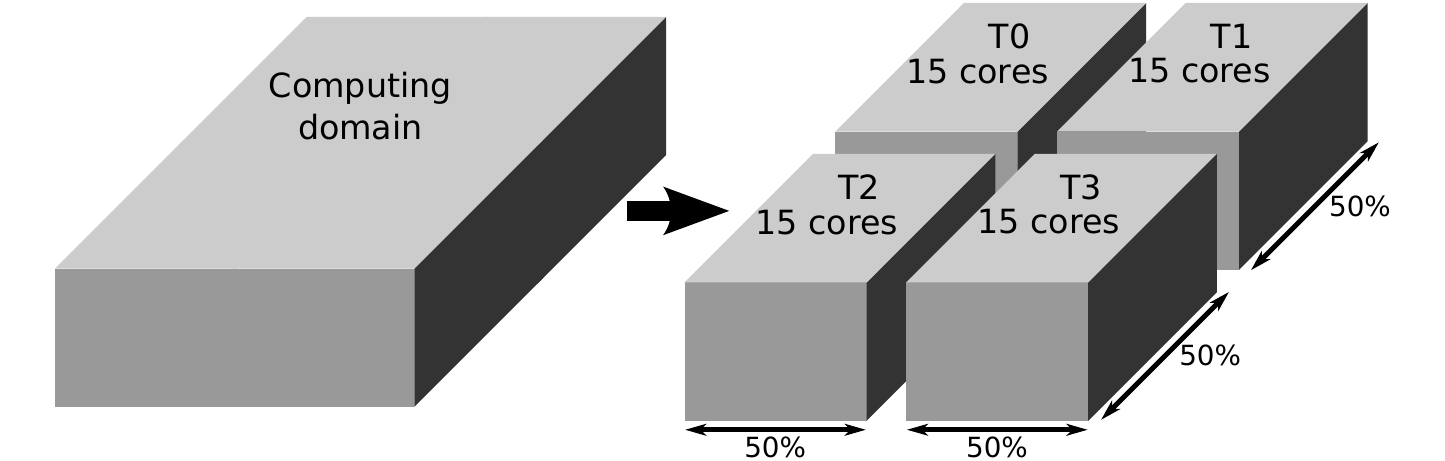}
\caption{Partitioning of Intel MIC's processing cores into 4 work teams}
\label{fig:MPDATAteam}
\end{center}
\end{figure}

The proposed decomposition contributes to the data traffic from the main memory to the cache hierarchy.
In consequence, a lot of inter- and intra-cache communications are generated between more than 200 Intel MIC's processing cores.
To improve the efficiency of the (3+1)D decomposition on Intel Xeon Phi, we provided partitioning of available cores (threads) into independent work teams.
As a result, the MPDATA computing domain is partitioned into $P$ sub-domains of different sizes, each of which is processed by a single work team of threads, according to the proposed (3+1)D decomposition.
The number of cores (threads) assigned to different teams can also be different.
Figure~\ref{fig:MPDATAteam} shows an example of partitioning of 60 Intel MIC's processing cores into $4$ teams, and partitioning of the MPDATA grid into $4$ pieces as well.

Within every time step, the work teams execute computations in parallel and independently of each other.
After each time step, the work teams are synchronized.
Each sub-domain is further partitioned into a number of MPDATA blocks, where subsequent blocks are processed one by one, and each block is processed in parallel by the corresponding work team.
Every block is further decomposed into sub-blocks, where each sub-block is processed by a certain thread of the work team.
A sequence of all the MPDATA stages is executed within every sub-block, taking into account the data dependencies.
This is achieved at the cost of some extra computations performed for halo regions by all teams.

Our best performance results on a single Intel Xeon Phi have been achieved so far by partitioning the 3D $n\times m\times l$ MPDATA domain in two dimensions $n$ and $m$ into four equal sub-domains, so that there is one-to-one mapping between these sub-domains and the teams of cores arranged in a $2\times2$ grid as illustrated in Figure~\ref{fig:MPDATAteam}. This partitioning allowed us to balance the load of the core teams and minimize the execution time of the application in comparison with all other partitioning shapes.

\subsection{Applying model-based partitioning algorithm to MPDATA decomposition}
\label{subsec:model}

Thus, the best load-balanced configuration of the MPDATA application on a Intel Xeon Phi arranges its cores in four $15$-core teams as shown in Figure~\ref{fig:MPDATAteam} and evenly partitions the $n\times m\times l$ computation domain between these teams, allocating a  $\frac{n}{2}\times\frac{m}{2}\times l$ sub-domain to each of the teams. In this subsection, we use the data-partitioning algorithm, presented in Section~\ref{sec:model}, to find a better partitioning of the computation domain between these teams of cores.  

As a first step, we build speed functions of the teams so that the speed of each team be represented by a function of problem size. In the case of MPDATA, the problem size is characterized by the size of the domain processed by the team and therefore represented by three parameters $n$, $m$ and $l$. In real-life NWP simulations $l$ is fixed \cite{SZU15}. Therefore, we build speeds of teams as functions of two parameters $n$ and $m$, setting $l$ to $128$, the value typically used in simulations. 

In general, the speed should be measured in equal-sized computation units performed per one time unit \cite{ijhpca2007}, for example, in flops. In the case of MPDATA, it is difficult to estimate the amount of arithmetic operations that will be executed during the processing of a $n\times m \times l$ computation domain. We know however that with a very high level of accuracy this amount is directly proportional to the number of cells in this domain. Therefore, we measure the speed in cells per second. 

The speed functions are built empirically by benchmarking the work teams for a range of problem sizes. For each problem size $(n,m)$, the speed is calculated as $\frac{n\times m\times 128}{t}$, where $t$ is the measured execution time. 

\begin{figure}[h!]
\begin{center}
\includegraphics[width=0.5\textwidth]{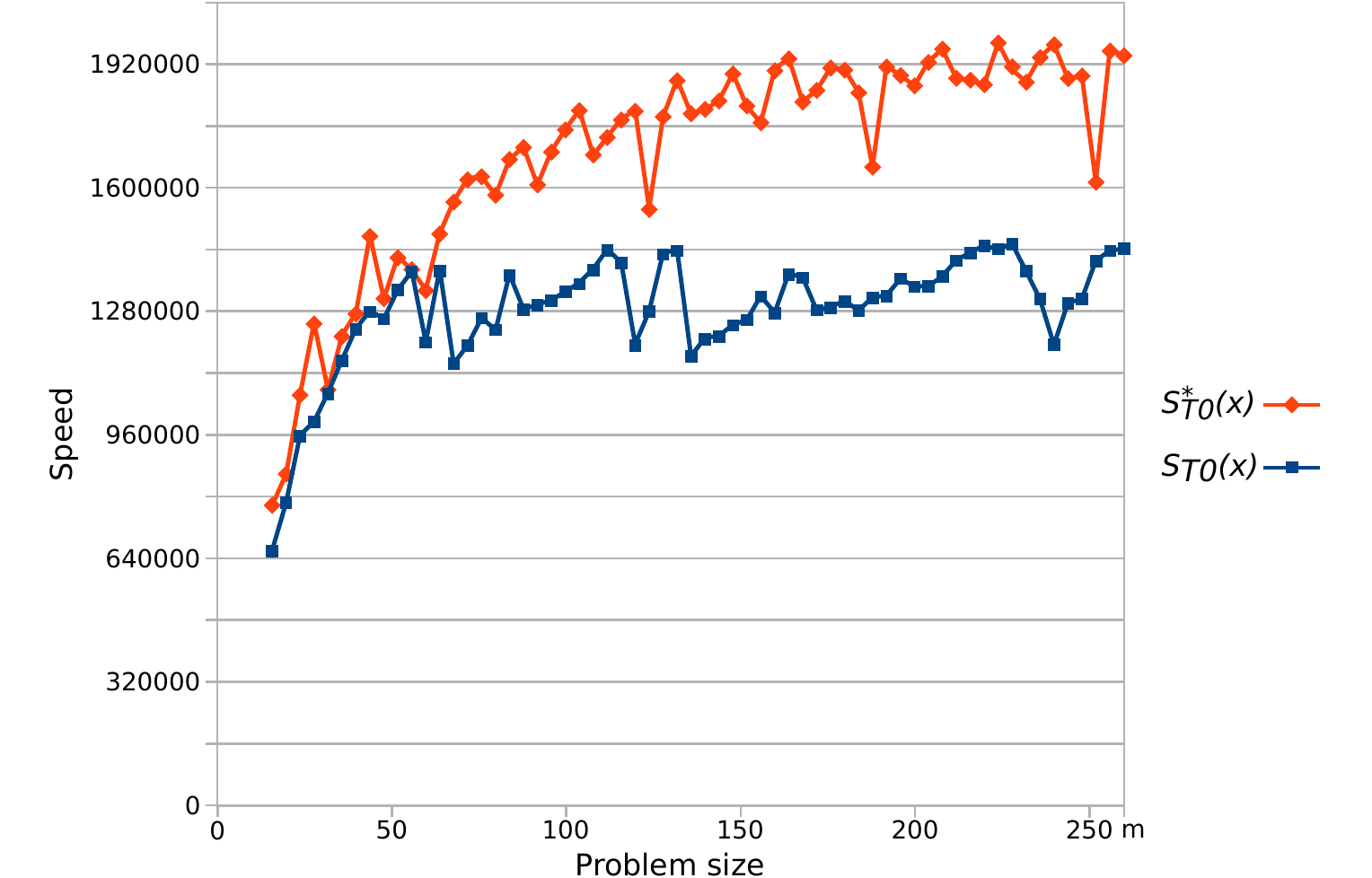}
\caption{Comparison of speed functions of team $T_0$, measured separately($S_{T_0}^*(x)$) and simultaneously with other three teams ($S_{T_0}(x)$) executing the same workload}
\label{fig:graph3_0}
\end{center}
\end{figure}

It has been shown \cite{ziming2011, zimingTC2014} that in modern multicore, manycore and hybrid platforms, where processing elements are coupled and share resources, the speed of one group of elements may depend on the load of others due to resource contention. Therefore, the groups cannot be considered as independent processing units and their speed cannot be measured separately and independently. In this work, we use the performance measurement method proposed in \cite{zimingTC2014} . According to this method, the performance of the four teams of cores is measured simultaneously rather than separately, thereby taking into account resource contention. To ensure the reliability of measurements, we repeat measurements multiple times. We only measure the computation time of every team without the overheads of inter-team synchronization required after each time step.  If the measurements were conducted separately, the measured performance of these teams would not reflect their actual performance during the execution of the application, and therefore performance optimization decisions based on the corresponding performance models would be inaccurate. Figure~\ref{fig:graph3_0} demonstrates the difference between the speed of team $T_0$ measured separately and simultaneously with other teams.

\begin{figure}[h!]
\begin{center}
\includegraphics[width=0.5\textwidth]{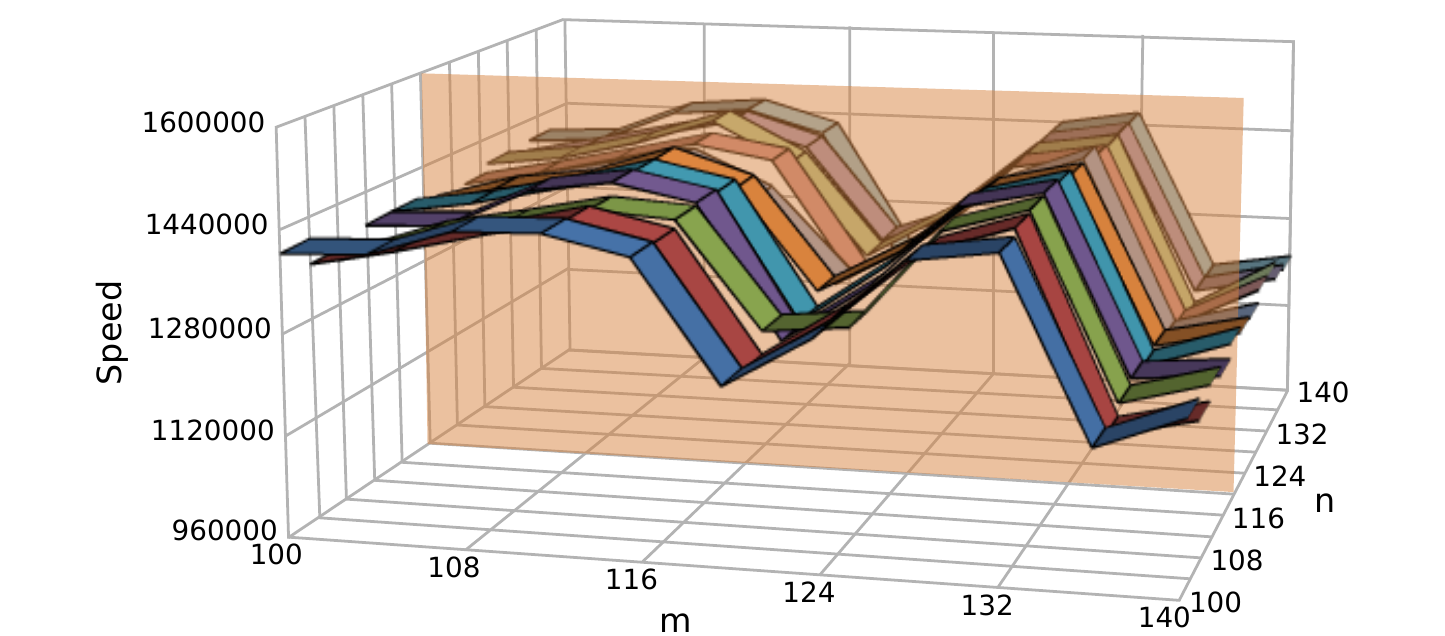}
\caption{Experimentally built speed of execution of the MPDATA workload by team $T_0$ as function of two parameters $n$ and $m$ ($l=128$)}
\label{fig:graph2}
\end{center}
\end{figure}

Figure~\ref{fig:graph2} illustrates the speed of team $T_0$ as a function of parameters $n$ and $m$ (remember that $l=128$).
The experimental points for the speed function were obtained with steps $\Delta n=\Delta m=4$ for both $n$ and $m$ parameters.

We can see that for a fixed value of $m$ the speed varies very slowly and very little with variation of $n$, staying nearly constant.
More detailed analysis of the speed functions confirms that the speed of team strongly depends on $m$ and very little depends on $n$. This observation allows us to assume that with a high level of accuracy the optimal (or at least a near optimal) partitioning of a $N\times M \times 128$  domain between the four teams can be obtained from the optimal even load-balanced partitioning, which allocates a sub-domain of size $\frac{N}{2} \times \frac{M}{2} \times 128$ to each team, by fixing the first parameter $n$ to $\frac{N}{2}$ and varying $m$. 

\begin{figure}[h!]
\begin{center}
\includegraphics[width=0.5\textwidth]{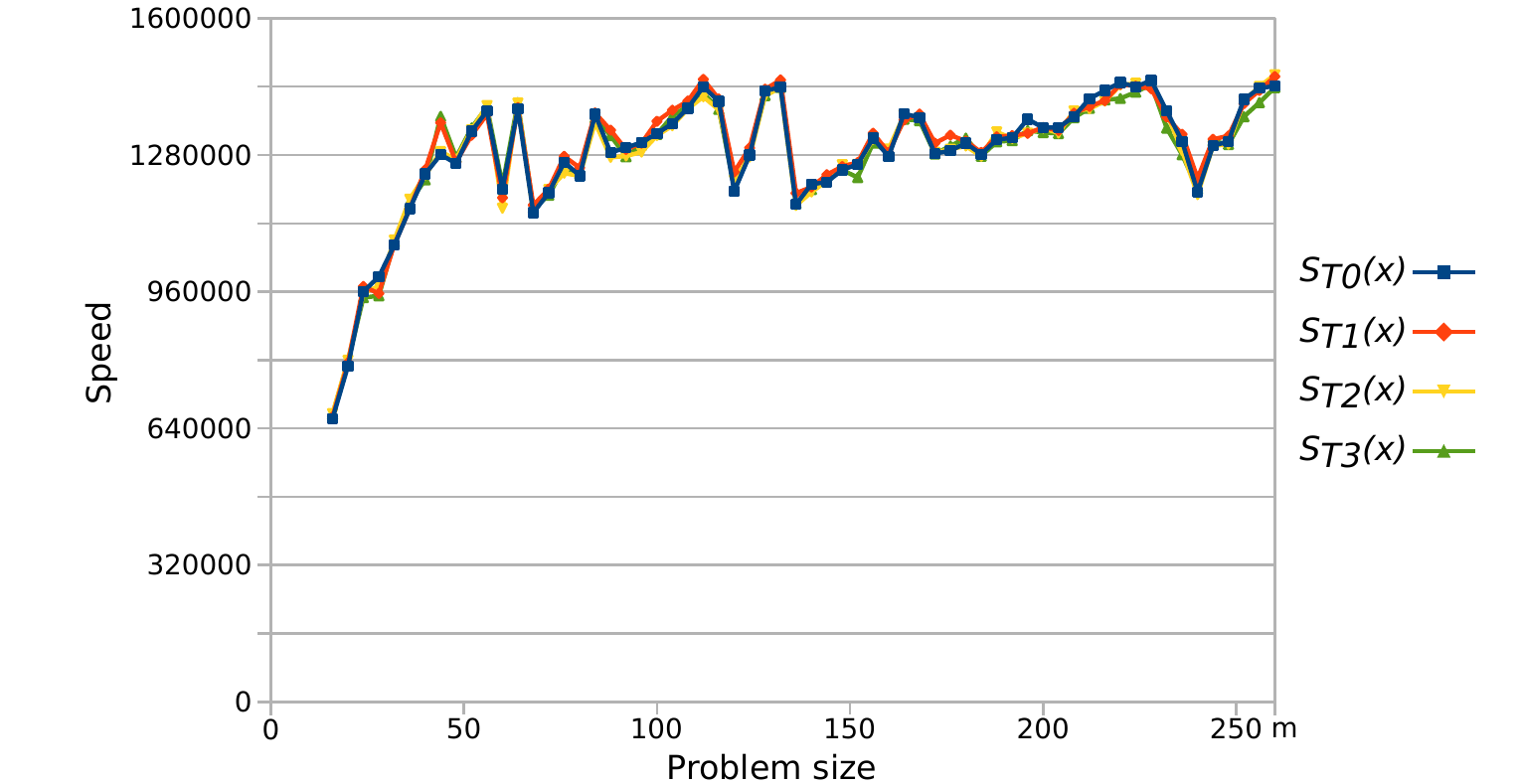}
\caption{Speeds of four teams built simultaneously as functions of parameter $m$ (\mbox{$n=120$} and \mbox{$l=128$})}
\label{fig:graph3}
\end{center}
\end{figure}

Mathematically, it means that we only have to deal with speed functions of just one parameter, $m$. These functions are obtained from the previously built speed functions of two parameters, $n$ and $m$, by fixing the parameter $n$. Geometrically, this can be illustrated as follows. The two-parameter speed functions are represented by surfaces. By fixing parameter $n$ to $\frac{N}{2}$, we cut the surfaces by a vertical plane $n=\frac{N}{2}$ as shown in Figure~\ref{fig:graph2}, obtaining on this plane curves, representing the one-parameter speed functions as shown in Figure~\ref{fig:graph3} for $n=120$.

\begin{figure}[h!]
\begin{center}
\includegraphics[width=0.5\textwidth]{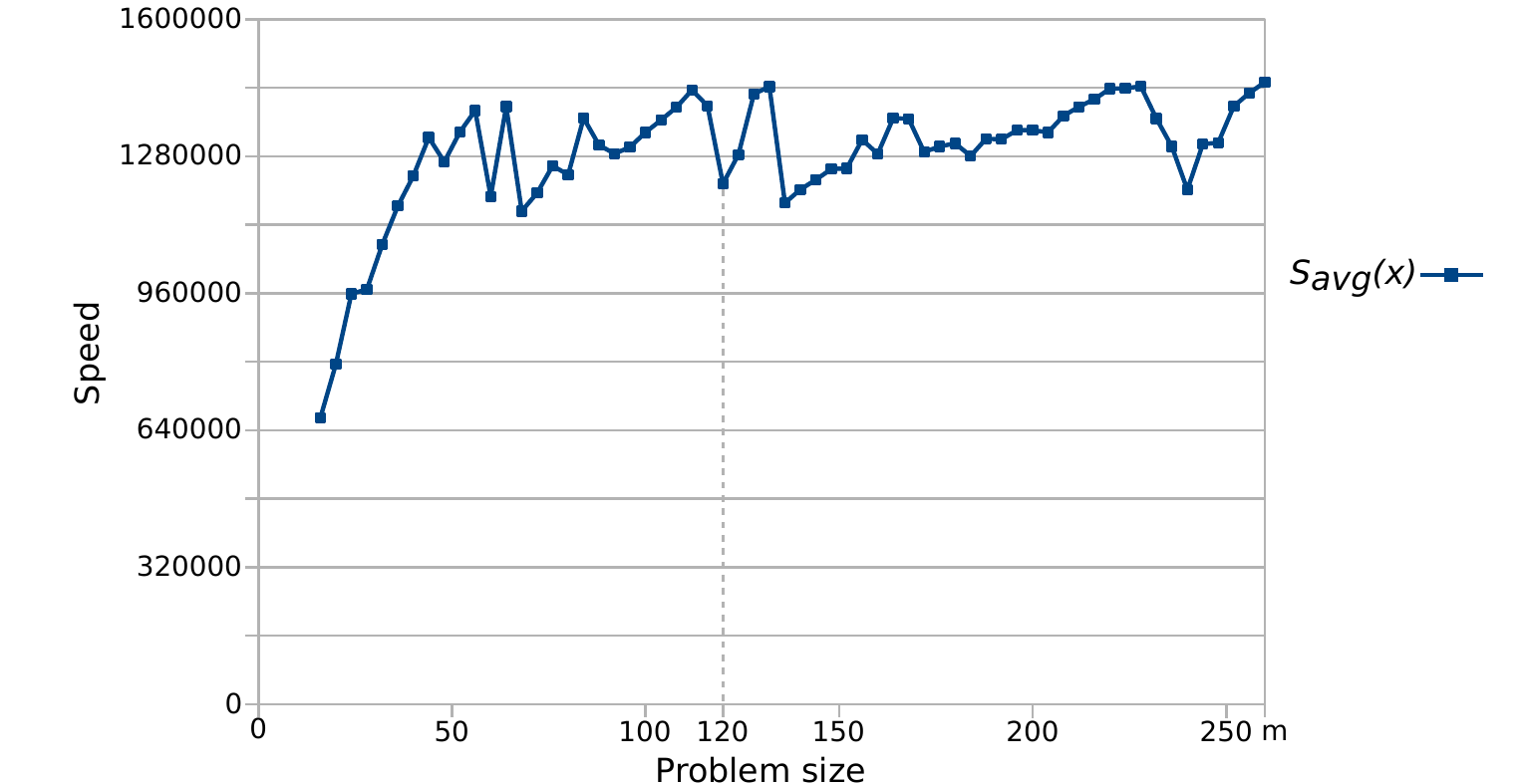}
\caption{Averaged speed of  teams built as a function of parameter $m$ (\mbox{$m=120$} and \mbox{$l=128$})}
\label{fig:graph4}
\end{center}
\end{figure}

Finally, as all four teams have very close speed functions (as can be seen in Figure~\ref{fig:graph3}), we calculate their average (shown in Figure~\ref{fig:graph4}) and use it as input to Algorithm~\ref{Alg:Homo} to find the optimal value of $m$ for each team.

More specifically, let the MPDATA domain be of size $240 \times 240 \times 128$. Then,
we consider our four teams as four identical abstract processors, $p=4$, the speed of each of which is given by the speed function shown in Figure~\ref{fig:graph4}. Note that in this function, the amount of workload is given in frames of cells of size $120 \times 128$, while the speed is given in cells per second. As pointed in Section~\ref{sec:model}, despite the unit of workload used to measure the speed (axis $y$) is $120 \times 128$ times greater than the unit of workload used to measure the size of workload (axis $x$), we can safely use this function as input to Algorthm~\ref{Alg:Homo}.

\begin{figure}[h!]
\begin{center}
\includegraphics[width=0.5\textwidth]{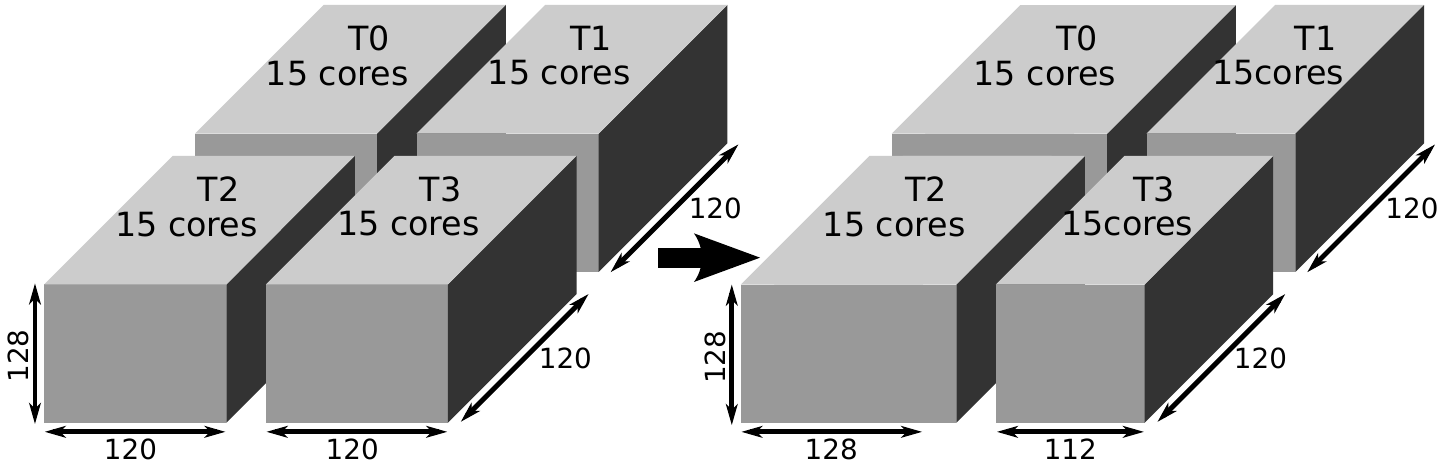}
\caption{Optimal partitioning of MPDATA of size $240 \times 240 \times 128$  between 4 teams}
\label{fig:TeamsNew}
\end{center}
\end{figure}

The solution returned by Algorthm~\ref{Alg:Homo} allocates $m=112$ frames to even abstract processors and $m=128$ frames to odd processors. This corresponds to partitioning of the $240 \times 240 \times 128$ domain into two sub-domains of size  $120 \times 112 \times 128$, allocated to teams $T_0$ and $T_2$, and two sub-domains of size  $120 \times 128 \times 128$, allocated to teams $T_1$ and $T_3$. The traditional load-balanced approach partitions the domain in four equal sub-domains of size $120 \times 120 \times 128$. This is illustrated in Figure~\ref{fig:TeamsNew}.


The theoretical execution time of the even partitioning is 1.486 seconds, while the theoretical execution time of the uneven partitioning returned by Algorithm~\ref{Alg:Homo} is 1.386:
\begin{equation*}
\begin{split}
t=\max(\frac{x_r}{s(x_r)},\frac{x_l}{s(x_l)})
\end{split}
\end{equation*}
\begin{equation*}
t= \max(\frac{120 \cdot 128 \cdot 128 }{1418579},\frac{120 \cdot 112 \cdot 128 }{ 1436742}) = 1.386[s].
\end{equation*}

\section{Experimental results}

\label{sec:result}

In this section, we experimentally evaluate the optimization technique presented in Section~\ref{sec:app}.


The performance results presented in this section are obtained for  double precision MPDATA computations corresponding to 40 time steps.
All the benchmarks are compiled as native executables using the Intel compiler (v.15.0.2), and run on the Intel Xeon Phi 7120P coprocessor.
To ensure the reliability of the results, measurements are repeated multiple times, and average execution times are used. 
We find the confidence interval and stop the measurements if the sample mean lies in the interval with the confidence level 95\%. 
We use Student's \textit{t}-test, assuming that the individual observations are independent and their population follows the normal distribution.

Table \ref{tab:240} includes both  theoretical and experimental execution times of MPDATA for  the domain of size $240 \times 240 \times 128$.
These results are obtained for different configurations of partitioning, including the traditional "load-balanced" partitioning ($\Delta m = 0$) and a range of "unbalanced" partitioning for different $\Delta m >0$.
The theoretically optimal $\Delta m$ returned by Algorithm~\ref{Alg:Homo} is equal to $8$, which corresponds to the configuration where each odd or even team processes the sub-domain of size $120\times128\times128$ or $120\times112\times128$ respectively.
In this case, the estimated execution time of 1.386 seconds is very close to the real computation time which is 1.364 seconds. 
According to experiments, the shortest execution time is achieved for $\Delta m=9$, when computations take 1.348 seconds. 

\begin{table}[h]
\caption{Theoretical and experimental execution times for MPDATA domain of size $240 \times 240 \times 128$ with different configurations of partitioning.
The odd work teams process the sub-domain of size $120 \times (120+\Delta m) \times 128$, while the even teams --$120 \times  (120-\Delta m) \times 128$.}
\label{tab:240}
\centering
\begin{tabular}{|c|c|c|c|}
\hline
Offset	&Theoretical	&	Experimental&		\\ 
$\Delta m$	&time [s]	&	time [s]&	Speedup	\\ \hline
0	&	1.486	&	1.548	&	1.000	\\	\hline
4	&	1.470	&	1.470	&	1.053	\\	\hline
6	&	1.401	&	1.374	&	1.127	\\	\hline
7	&	1.422	&	1.361	&	1.137	\\	\hline
8	&	1.386	&	1.364	&	1.135	\\	\hline
9	&	1.398	&	1.348	&	1.148	\\	\hline
10	&	1.397	&	1.352	&	1.145	\\	\hline
11	&	1.429	&	1.372	&	1.129	\\	\hline
12	&	1.402	&	1.368	&	1.131	\\	\hline
\end{tabular}
\end{table}

Comparing the experimental and theoretical times, we can see that 
the accuracy of theoretical prediction is very good, with prediction errors being as small as $2-4\%$. 
In general, we can identify 
two
main factors contributing into the prediction error:
\begin{itemize}
\item While the experimentally built speed functions of teams $T_0$, $T_1$, $T_2$ and $T_3$ are not identical, suggesting some degree of their heterogeneity in execution of the MPDATA workload, our theoretical model  considers them homogeneous and represents their speed by the average of the real speed functions, which is then used as input to Algorithm~\ref{Alg:Homo}.
\item 
During the construction of the speed functions, the speed of a team for problem size $n\times m\times l$ is measured when other teams process in parallel sub-domains of the same size, $n\times m\times l$.  However, during the execution of the application in our experiments different teams process sub-domains of slightly different sizes when $\Delta m\neq 0$.
\end{itemize}

Table \ref{tab:240} also demonstrates  the performance gain from applying the proposed load-imbalancing optimization.
For the imbalanced configurations presented in this table, we notice a better performance than for the load-balanced configuration of the MPDATA decomposition.
The largest performance gain is achieved for $\Delta m=9$, giving the speedup of 1.148x.

\begin{table}[h]
\caption{Experimental time for all work teams with different partitionings: 
the odd work teams process the sub-domain of size $120 \times (120+\Delta m) \times 128$, while the even teams --$120 \times  (120-\Delta m) \times 128$.}
\label{tab:teams}
\centering
\begin{tabular}{|c||c|c|c|c||c|}
\hline
Offset	&	\multicolumn{5}{|c|}{ Experimental time [s]}		\\ \cline{2-6}
$\Delta m$	&	Team 0	&	Team 1	&	Team 2	&	Team 3	&	Total	\\ \hline
0	&	1.515	&	1.498	&	1.518	&	1.503	&	1.548	\\	\hline
4	&	1.456	&	1.247	&	1.455	&	1.249	&	1.470	\\	\hline
6	&	1.364	&	1.161	&	1.359	&	1.162	&	1.374	\\	\hline
7	&	1.355	&	1.161	&	1.341	&	1.168	&	1.361	\\	\hline
8	&	1.355	&	1.166	&	1.349	&	1.172	&	1.364	\\	\hline
9	&	1.340	&	1.155	&	1.335	&	1.161	&	1.348	\\	\hline
10	&	1.345	&	1.141	&	1.337	&	1.152	&	1.352	\\	\hline
11	&	1.363	&	1.156	&	1.357	&	1.154	&	1.372	\\	\hline
12	&	1.360	&	1.163	&	1.353	&	1.165	&	1.368	\\	\hline
\end{tabular}
\end{table}

Table \ref{tab:teams} complements the results in Table \ref{tab:240} giving experimental execution times of the individual teams.
We can clearly see a significant difference between the execution times measured for the odd and even teams when $\Delta m \ne 0$.
Obviously, this difference is caused by the unbalanced workloads for the odd and even teams.
However, the total execution time is shorter than in the case of balanced workloads ($\Delta m=0$).

\begin{figure}[h!]
\begin{center}
\includegraphics[width=0.5\textwidth]{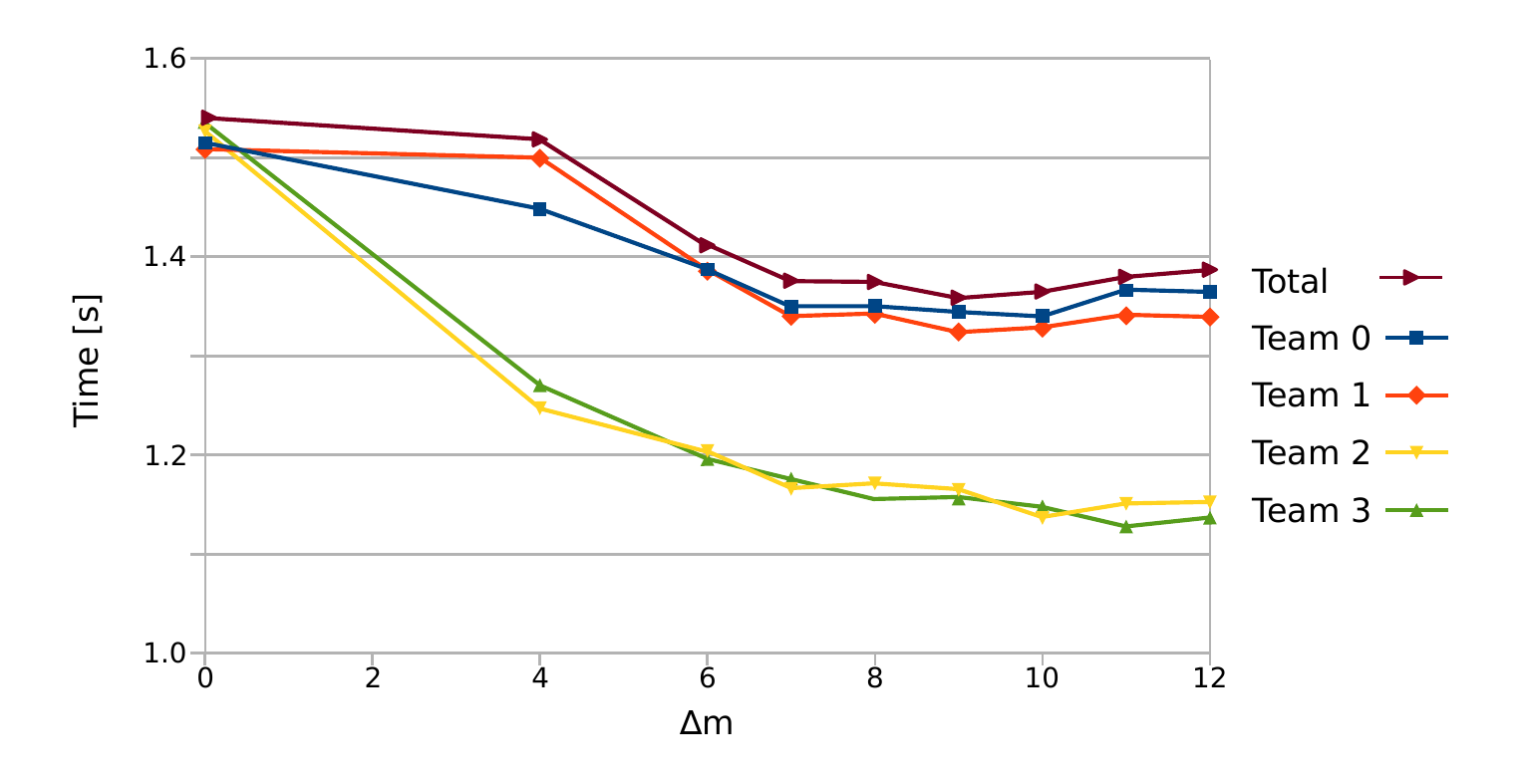}
\caption{Experimental execution times measured for individual work teams and the total execution time measured for the whole MPDATA workload}
\label{fig:result}
\end{center}
\end{figure}

Table \ref{tab:teams} also shows that the total execution time is always slightly longer than the maximum time among all teams. 
It is mainly due to the fact that the computation time of every team is measured without the overheads of inter-team synchronization required after each time step.
In addition, the results in Table \ref{tab:teams} are presented in \mbox{Figure~\ref{fig:result}} in a graphical form.

Finally, we evaluate the proposed model-based partitioning algorithm for the MPDATA domain of size $480 \times 480 \times 128$.
As in the previous case, the application is executed for different configurations of partitioning, for a range of $\Delta m$.
In this case, however, the theoretically optimal configuration returned by Algorithm~\ref{Alg:Homo} is exactly  the same as the experimentally optimal one, both achieved when $\Delta m=20$. 
The prediction errors are also smaller in this case, not exceeding $3\%$. 
The experimental execution time for $\Delta m=20$ is 5.338 seconds, in comparison with 6.140 seconds for the even partitioning.This allows us to accelerate the MPDATA computations by 1.15 times.
Moreover, the performance gain is also observed for other unbalanced configurations, but it is smaller than 1.15x. 
The results of these experiments are included in Table \ref{tab:480}.

\begin{table}[h]
\caption{Theoretical and experimental execution times for MPDATA domain of size $480 \times 480 \times 128$ with different configurations of partitioning.
The odd work teams process the sub-domain of size $240 \times (240+\Delta m) \times 128$, while the even teams -- $240 \times  (240-\Delta m) \times 128$.}
\label{tab:480}
\centering
\begin{tabular}{|c|c|c|c|}
\hline
Offset	&Theoretical	&	Experimental&		\\ 
$\Delta m$	&time [s]	&	time [s]&	Speedup	\\ \hline 
0	&	6.136	&	6.140	&	1.000	\\ \hline
4	&	5.731	&	5.681	&	1.081	\\ \hline
8	&	5.809	&	5.806	&	1.058	\\ \hline
12	&	5.543	&	5.453	&	1.126	\\ \hline
16	&	5.509	&	5.418	&	1.133	\\ \hline
20	&	5.499	&	5.338	&	1.150	\\ \hline
24	&	5.624	&	5.477	&	1.121	\\ \hline
\end{tabular}
\end{table}

\section{Conclusion}
\label{sec:con}
Modern compute nodes are characterized by both the increasing number of (possibly, heterogeneous) processing elements and a high level of complexity of their integration. Various resources such as caches and data links are shared in an hierarchical and non-uniform way. This makes the development of efficient applications for such platforms a very difficult and challenging task. It would be naive to expect that the performance profile of real-life scientific applications on these platforms will always be comfortably nice and smooth to suit traditional load-balancing techniques used for minimization of their computation time. Therefore, new optimization approaches that do not rely on such increasingly unrealistic assumptions are needed.
This work has presented one such approach and demonstrated its applicability to optimization of a real-life application on a modern HPC platform.
 
\textbf{Acknowledgments} \\ This research was conducted with the financial support of NCN under grant no. UMO-2011/03/B/ST6/03500.
We gratefully acknowledge the help and support provided by Jamie Wilcox from Intel EMEA Technical Marketing HPC Lab.
This work is partially supported by EU under the COST Program Action IC1305: Network for Sustainable Ultrascale Computing (NESUS). 

\bibliographystyle{abbrv}
\bibliography{Lastovetsky_Szustak_sc15}

\end{document}